\documentclass[letters,a4paper,fleqn,usenatbib]{mnras}

\usepackage{newtxtext,newtxmath}

\usepackage[T1]{fontenc}
\usepackage{ae,aecompl}

\usepackage{graphicx}	
\usepackage{amsmath}	
\usepackage{amssymb}	

\title[Gamma radiation from the vicinity of black holes]
{Enhanced gamma radiation toward the rotation axis
	from the immediate vicinity of extremely rotating black holes}

\author[Y. Song et al.]{
	Yoogeun Song,$^{1,2}$\thanks{E-mail : ygsong1004@gmail.com}
Hung-Yi Pu,$^{3}$
Kouichi Hirotani,$^{3}$\thanks{E-mail : hirotani@tiara.sinica.edu.tw}
Satoki Matsushita,$^{3}$ \and
Albert K. H. Kong,$^{4}$ and
Hsiang-Kuang Chang$^{4}$  \\
$^{1}$ Korea Astronomy and Space Science Institute, Daejeon 305-348, Republic of Korea\\
$^{2}$ University of Science and Technology,  Daejeon 305-350, Republic of Korea \\
$^{3}$ Academia Sinica, Institute of Astronomy and Astrophysics (ASIAA),
PO Box 23-141, Taipei, Taiwan\\
$^{4}$ Institute of Astronomy, Department of Physics, 
       National Tsing Hua University,
       No. 101, Section 2, Kuang-Fu Road, Hsinchu,\\ 30013, Taiwan
}

\date{Accepted XXX. Received YYY; in original form ZZZ}

\pubyear{2017}

\def\beq{\begin{equation}}
\def\eeq{\end{equation}}
\def\bea{\begin{eqnarray}}
\def\eea{\end{eqnarray}}

\def\nn{\nonumber}

\def\vphi{\varphi}

\def\rrF{{\rm F}}

\def\rrB{{\rm B}}
\def\rrH{{\rm H}}

\def\OmegaF{\Omega_\rrF}

\def\ra{\rightarrow}

\begin{document}
\label{firstpage}
\pagerange{\pageref{firstpage}--\pageref{lastpage}}
\maketitle


\begin{abstract}
	We investigate the acceleration of electrons and positrons
	by magnetic-field-aligned electric fields in the polar funnel of an accreting black hole (BH).
	Applying the pulsar outer-gap theory to BH magnetospheres,
	we find that such a lepton accelerator arises
	in the immediate vicinity of the event horizon due to frame-dragging,
	and that their gamma-ray luminosity increases
	with decreasing accretion rate.
	Furthermore, we demonstrate that
	the gamma-ray flux is enhanced along the rotation axis
	by more than an order of magnitude if the BH spin increases
	from $a=0.90M$ to $a=0.9999M$.
	As a result, if a ten-solar-mass, almost-maximally rotating BH
	is located within 3~kpc,
	when its accretion rate is between 0.005\% and 0.01\% 
	of the Eddington rate,
	its high-energy flare becomes detectable with
	the Fermi/Large Area Telescope,
	provided that the flare lasts longer than 1.2 months
	and that we view the source nearly along the rotation axis.
	In addition, its very-high-energy flux is marginally detectable with
	the Cherenkov Telescope Array,
	provided that the flare lasts longer than a night
	and that our viewing angle is about 45 degrees
	with respect to the rotation axis.

\end{abstract}





\section{INTRODUCTION}

It is widely accepted that the electromagnetic extraction
of the rotational energy of black holes 
\citep{bla77} 
is one of the promising mechanisms for powering galactic 
black hole (BH) binaries and active galactic nuclei.
The extracted energy is transported outward
in the form of the Poynting and the plasmas' kinetic energy fluxes,
and finally dissipated at large distances.
However, a portion of this energy can also be dissipated 
 in the vicinity of the BH,
if there appears a charge deficit in the magnetosphere
in the same manner as in pulsar emission models
\citep{stur71,harding78,cheng86a,romani96,hiro13,taka16}.

In this context, \citet{bes92} applied the pulsar outer-gap model
to BH magnetospheres and pointed out the possibility of 
an efficient pair-production cascade near the null-charge surface,
where the Goldreich-Julian (GJ) charge density vanishes
due to the space-time frame-dragging around a rotating BH.
Then \citet{hiro98}  demonstrated that the stationary numerical solution
obtained from the set of Maxwell-Boltzmann equations
is consistent with the gap closure condition
\citep[][hereafter H17]{hiro17}. 


In order to consider bright gap emissions, 
\citet{nero07,levi11,brod15}, and \citet[][hereafter HP16]{hiro16}
examined spatially extended gaps, 
which can be possible in a weak soft photon field.
For example, when the mass accretion rate is typically less than
1\% of the Eddington rate, the accreting plasmas form 
a radiatively inefficient accretion flow (RIAF),
emitting radio to infrared (i.e., soft) photons via synchrotron process
and MeV photons via free-free and inverse-Compton (IC) processes
\citep{ichimaru77,narayan94,abram95,mahad97,manmoto00}.
Under such a low accretion environment,
gap-emitted TeV photons do not efficiently collide 
with the soft photons,
leading to an un-screened, spatially extended gap, 
which has a much greater electric potential drop
than denser soft-photon-field cases.

More recently, \citet[][hereafter H16]{hiro16b} quantified the BH gap model
in the two-dimensional poloidal plane,
solving the set of the inhomogeneous part of the Maxwell equations,
the motion of the created electrons and positrons 
(which are referred to as leptons in this letter) within the gap, and
the radiative transfer equation for the gap-emitted photons,
assuming a mono-energetic approximation for 
the lepton distribution functions.
Then H17 
considered an inhomogeneous soft photon field
to examine the $\gamma$-ray emission properties of 
super-massive BHs (SMBHs),
solving the distribution functions of the accelerated leptons
explicitly from their Boltzmann equations.
In these two works, however, the radial magnetic field strength,
$B^r$, is assumed to be uniform on the event horizon, 
because they assumed only moderate BH spins, namely
$a \le 0.90M$, where $M$ is the BH's mass.

However, it was suggested by numerical simulations 
that the magnetic flux concentrates toward the rotation axis
as the BH extremely rotates, $a \rightarrow M$
\citep{komiss07,tchek10}.
 Indeed, it causes a significant impact on the BH-gap emission.
In the present paper, we therefore investigate extremely rotating BHs
and demonstrate that their gap emission is strongly beamed toward the
rotation axis as $a \ra M$.

After briefly describing our methodology in \S\ref{sec:model},
we present our results in \S\ref{sec:results},
and discuss the enhancements from  previous works in \S\ref{sec:disc}.

\section[]{Black hole gap model}
\label{sec:model}
%
In this letter, we follow the methodology presented in H17,
which is briefly described in 
\S\ref{sec:geometry}--\ref{sec:BDCs},
incorporating the effect of the magnetic field lines
rearranging laterally and concentrating around the axis of rotation
(\S\ref{sec:poleward}).

\subsection[]{Background geometry}
\label{sec:geometry}

Around a rotating BH, 
the background geometry is described by the Kerr metric
\citep{kerr63}.
In the Boyer-Lindquist coordinates \citep{boyer67}, 
adopting the geometrized unit, $c=G=1$, where
$c$ and $G$ denote the speed of light and the gravitational constant, respectively,
we obtain the following line element,
\begin{equation}
ds^2= g_{tt} dt^2
+2g_{t\varphi} dt d\varphi
+g_{\varphi\varphi} d\varphi^2
+g_{rr} dr^2
+g_{\theta\theta} d\theta^2,
\label{eq:metric}
\end{equation}
where 
\begin{equation}
g_{tt} 
\equiv 
-\frac{\Delta-a^2\sin^2\theta}{\Sigma},
\qquad
g_{t\varphi}
\equiv 
-\frac{2Mar \sin^2\theta}{\Sigma}, 
\label{eq:metric_2}
\end{equation}
\begin{equation}
g_{\varphi\varphi}
\equiv 
\frac{A \sin^2\theta}{\Sigma} , 
\qquad
g_{rr}
\equiv 
\frac{\Sigma}{\Delta} , 
\qquad
g_{\theta\theta}
\equiv 
\Sigma;
\label{eq:metric_3}
\end{equation}
\bea
&&\Delta \equiv r^2-2Mr+a^2, ~~~~~~~
\Sigma\equiv r^2 +a^2\cos^2\theta, \nn\\
&&A \equiv (r^2+a^2)^2-\Delta a^2\sin^2\theta.
\eea
The horizon radius, $r_{\rm H} \equiv M+\sqrt{M^2-a^2}$,
is obtained by $\Delta=0$, 
where $M$ corresponds to the gravitational radius.
The spin parameter becomes $a=M$ for a maximally rotating BH,
and $a=0$ for a non-rotating BH. 

We assume that the non-corotational potential $\Phi$
depends on $t$ and $\varphi$ only through
the form $\varphi-\Omega_{\rm F} t$, where
$\Omega_{\rm F}$ denotes the magnetic-field-line
rotational angular frequency.
Under this ``stationary'' approximation,
Gauss's law gives us the Poisson equation
that describes $\Phi$ in a three dimensional magnetosphere
(eq.~[15] of H16),
\begin{equation}
-\frac{1}{\sqrt{-g}}
\partial_\mu 
\left( \frac{\sqrt{-g}}{\rho_{\rm w}^2}
g^{\mu\nu} g_{\varphi\varphi}
\partial_\nu \Phi
\right)
= 4\pi(\rho-\rho_{{\rm GJ}}),
\label{eq:pois}
\end{equation}
where the GR GJ charge density
is defined as
\citep{GJ69,mestel71,hiro06}
\begin{equation}
\rho_{{\rm GJ}} \equiv 
\frac{1}{4\pi\sqrt{-g}}
\partial_\mu \left[ \frac{\sqrt{-g}}{\rho_{\rm w}^2}
g^{\mu\nu} g_{\varphi\varphi}
(\Omega_{\rm F}-\omega) F_{\varphi\nu}
\right],
\label{eq:def_GJ}
\end{equation}
where $\omega \equiv - g_{t \vphi} / g_{\vphi\vphi}$ denotes the frame-dragging frequency at each point.
If the real charge density $\rho$ deviates from the
rotationally induced GJ charge density,
$\rho_{\rm GJ}$, in some region,
equation~(\ref{eq:pois}) shows that
$\Phi$ changes as a function of position.
Therefore, an acceleration electric field, 
$E_\parallel= -\partial \Phi / \partial s$,
arises along the magnetic field line,
where $s$ denotes the distance along the magnetic field line.
A gap is defined as the spatial region in which $E_\parallel$
is non-vanishing,
and appears around the null-charge surface,
where $\rho_{{\rm GJ}}$ changes sign (\S2.3.2 of HP~16).
Because of the frame-dragging effect,
the null-charge surface (and hence a gap) appears
near the horizon (Fig.~\ref{fig:sideview}).

\subsection[]{Radiatively inefficient accretion flow}
\label{sec:RIAF}
In order to quantify the gap electrodynamics, 
we need to compute the pair creation rate. 
In the same way as HP16, H16, and H17,
we consider only low-accretion-rate cases and adopt 
the advection-dominated-accretion flow (ADAF) solution 
that was obtained analytically by \citet{mahad97}
as a RIAF.
We compute the propagation of the ADAF-emitted soft 
photons (from radio to soft-$\gamma$-ray energies)
in the 3-D rotating magnetosphere,
and tabulate their specific intensity at each point in the 2-D
poloidal plane, assuming axisymmetry.
For details, see \S3 of H17.
The tabulated specific intensity is used to compute the photon-photon
collisions and IC scatterings
inside and outside the gap,
whose electrodynamics is outlined in the next subsection.

\subsection[]{Gap electrodynamics}
\label{sec:electrodynamics}
In the same way (that quantities do not depend on $\vphi - \OmegaF t$)  as HP17,
we solve the stationary gap solution
from the set of the Poisson equation for $\Phi$,
the equations of motion for the created leptons,
and the radiative transfer equation for the emitted photons.
Note that the non-uniform, inhomogeneous ADAF photon field
at each position in the BH magnetosphere is explicitly
taken into account (\S2 and Fig.~2 of H17).
For the formalism, readers may refer to \S4 of H17, for details.
The improvement over H17 is that we now take into account
the poleward concentration of the magnetic field lines
as $a \rightarrow M$  (with BH mass $M$, not gravitational radius)
\citep[see Fig.~7 of][]{tchek10},
instead of assuming a constant $B^r$ on $\theta$.

\subsection[]{Boundary conditions}
\label{sec:BDCs}
We impose the same boundary conditions as H17;
namely, electrons, positrons, and $\gamma$-rays 
do not penetrate into the gap,
except for the ADAF-emitted photons.
We assume axisymmetry and solve the gap on the 2-D poloidal plane.
We also impose a reflection symmetry with respect to the rotation axis,
$\theta=0$,
and assume that the gap is bounded from the equatorial, accreting
region at $\theta=60^\circ$.
Both the inner (i.e., BH-ward) and outer boundaries are solved
as free boundaries. 
Their positions are determined by the following two conditions:
(1) the created electric current within the gap is externally given, and
(2) a stationary gap solution is obtained by the gap closure condition
(\S4.2.5 of H17).

\subsection[]{Poleward concentration of magnetic field}
\label{sec:poleward}
In HP16 and H16, we computed the radial component of the magnetic field strength as
\beq
B^r = B_\rrH  (2M/r)^2,
\eeq
 with a dimensionless accretion rate $\dot{m}$, where
\beq
B_\rrH = 4 \times 10^8 \sqrt{\frac{\dot{m}}{M_1}}~ G
\eeq
denotes the equipartition magnetic field strength at $r = 2M$ \citep{levi11}, and $M_1 \equiv M/(10^1 M_\odot)$ for the BH's mass $M$.
Here, $M_\odot$ denotes the solar mass.
In the present paper, in order to incorporate the $\theta$ dependence, we compute $B^r$ with a function $f_{\rm B}$ as
\begin{equation}
B^r= f_{\rm B} (\theta; a) \frac{\partial_\theta A_\varphi ^{(0)}}{\Sigma\sin\theta},
\label{eq:Br}
\end{equation}
where 
\beq
A_\varphi ^{(0)} \equiv - (2 M)^2 B_\rrH \cos \theta ~.
\eeq
If we set $f_\rrB = 1$, we obtain  $B^r = B_\rrH  (2M/r)^2$ for $r \gg 2M$.
Note that the geometrical factor $g_{tt}+g_{t\varphi}\Omega_{\rm F}$
is not multiplied in the right-hand side
\citep{came86a,came86b}
to avoid a sign reversal of $B^r$ near the horizon.

Examining magnetohydrodynamic simulations,
\citet{tchek10} demonstrated that
the magnetic field lines progressively bunch up toward the rotation axis
as the BH spin increases.
For instance, if $a=0.5$, $f_{\rm B}$ takes an almost constant value 
of $0.29$ for $0^\circ < \theta < 90^\circ$.
However, when $a=0.9$, $f_{\rm B}$ becomes
$0.60$, $0.58$, $0.57$, $0.55$, $0.52$, $0.46$, $0.37$, and $0.37$
at
$\theta=0^\circ$, $5.3^\circ$, $10.8^\circ$, $16.6^\circ$, 
$23.6^\circ$, $34.3^\circ$, $63.0^\circ$, and $90^\circ$, respectively.
In the extreme case of $a=0.9999M$, they obtained
$2.87$, $2.62$, $2.12$, $1.62$, $1.25$, $0.87$, $0.62$, and $0.40$
at the same $\theta$'s.
Therefore, at the rotation axis,
$B^r$ becomes $2.1$ times stronger 
if $a$ increases from $0.50M$ to $0.90M$,
and becomes $9.9$ times stronger 
if it increases to $0.9999M$.
However, at a middle latitude, $\theta=23.6^\circ$,
$B^r$ becomes only $1.8$ to $4.3$ times stronger for the same values of $a$.
This poleward concentration of the magnetic fluxes
causes a significant increase of the gap emission 
along the rotation axis,
as will be described in the next section.

\section{Results}
\label{sec:results}
\setcounter{equation}{0}
\renewcommand{\theequation}{\arabic{section}.\arabic{equation}}

%
In the present paper, we adopt $M=10 M_\odot$ (i.e., stellar-mass) as the BH mass, 
and compare the results for two different spins, 
$a=0.9999M$ and $a=0.90M$.

\subsection{The Magnetic-Field-Aligned Electric Field}
\begin{figure}
	\vspace{-7.0 truecm}
	\includegraphics[width=8.0cm]{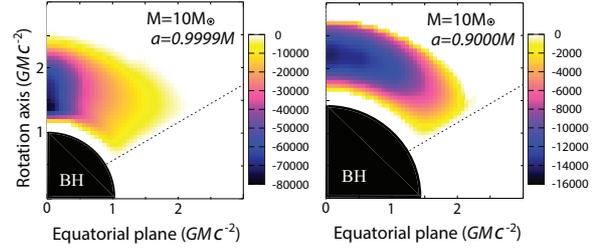}
	\caption{
		\small
		Side view of the magnetic-field-aligned electric field,
		$E_\parallel$ (in statvolt cm${}^{-1}$),
		exerted in the immediate vicinity of the event horizon.
		The black hole (BH, the filled circle on the bottom left corner)
		rotates around the ordinate.
		Both axes are in the Boyer-Lindquist coordinate
		and normalized by the gravitational radius of the BH.
		The thin black dotted line shows the boundary between 
		the polar funnel in which we solve the gap
		and the equatorial region
		where the ADAF photons are emitted and illuminates the polar funnel.
		The dimensionless accretion rate is set to be
		$\dot{m}=1.77 \times 10^{-4}$.
		The left panel is for an extremely rotating case
		with a spin parameter $a=0.9999M$, while
		the right panel for a rapidly rotating case with $a=0.90M$.
		It is clear that the gap position can be solved under a free boundary problem (\S\ref{sec:BDCs}).
	}
	\label{fig:sideview}
\end{figure}

We begin by considering the distribution of 
acceleration electric field.
In Fig.~\ref{fig:sideview}, we present $E_\parallel(r,\theta)$ 
for $a = 0.9999 M$ and $a = 0.90 M$.
In both panels, the peak of $\vert E_\parallel \vert$ 
appears at the pole, $\theta=0^\circ$.
Their maximum values attain
$7.91 \times 10^4 \mbox{statvolt cm}^{-1}$ for $a=0.9999M$, and
$1.42 \times 10^4 \mbox{statvolt cm}^{-1}$ for $a=0.90M$.
Thus, $E_\parallel$ increases about five times 
from the case of $a=0.90M$ to the case of $a=0.9999M$,
which is consistent with the increase of 
$f_{\rm B}$ from $0.60$ (for $a=0.90M$) to $2.87$ (for $a=0.9999M$)
at $\theta=0$.
This enhancement of $\vert E_\parallel \vert$ 
is due to the poleward concentration  
of the magnetic field lines as $a \rightarrow M$ \citep{tchek10}.

\subsection{Gap Emission versus Colatitudes}
\begin{figure}
			\vspace{-0.5 truecm}
	\includegraphics[width=8.0truecm,angle=0]{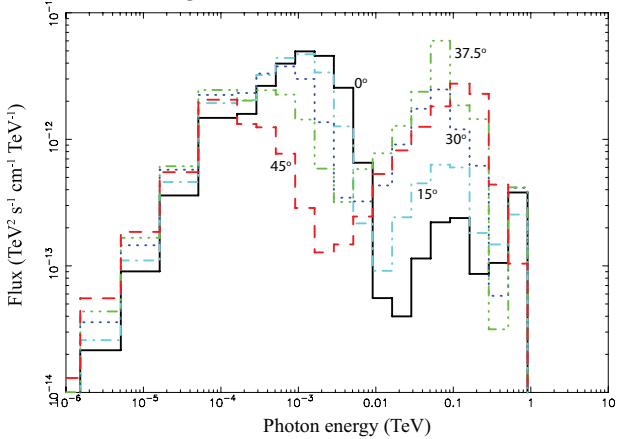}
	\caption{
		\small
		Spectral energy distribution (SED) of 
		the gap emission for a stellar-mass BH with 
		$M=10 M_\odot$ and $a=0.9999M$.
		The distance is assumed to be 3~kpc.
		The dimensionless accretion rate is chosen to be
		$\dot{m}=7.49 \times 10^{-5}$.
		The black solid, cyan dot-dashed, blue dotted,
		green dash-dot-dot-dotted, and red dashed lines
		show the SED viewed at the colatitudes
		$\theta=0^\circ$, $15^\circ$, $30^\circ$, $37.5^\circ$, and $45^\circ$,
		respectively. 
	}
	\label{fig:SED_1e1_a9999_5th}
\end{figure}

Poleward enhancement of $\vert E_\parallel \vert$ as $a \rightarrow M$
results in a poleward enhancement of the outward emission from the gap.
In Fig.~\ref{fig:SED_1e1_a9999_5th},
we compare the resultant $\gamma$-ray spectra 
at five discrete colatitudes, 
for an extremely rotating case, $a=0.9999M$.
The spectra peak in two different $\gamma$-ray energies.
The lower-energy peak appears 
in high-energy (HE) $\gamma$-rays,
between 0.05~GeV and 10~GeV,
while the higher-energy peak appears
in very-high-energy (VHE) $\gamma$-rays,
between 0.05~TeV and 1~TeV.
The HE emission is due to the curvature process 
while the VHE one is due to the IC process.
It follows that the HE emission becomes most luminous and hardest
along the rotation axis (as the solid line shows)
and that the VHE emission becomes more luminous 
and slightly harder along middle latitudes
(as the dotted, dash-dot-dot-dotted, and dashed lines show).
Since the curvature process dominates the IC process
for stellar-mass BHs,
the total energy flux integrated between MeV and TeV energies
is enhanced toward the rotation axis
for such an extremely rotating BH as $a=0.9999M$.
However, such an enhancement is modest for $a=0.90M$, 
as can be understood from the moderate enhancement of $B^r$
toward $\theta=0^\circ$ for $a=0.90M$ (\S\ref{sec:poleward}).


\subsection{Gap emission versus accretion rate}
\begin{figure}
	\includegraphics[width=8.0truecm,angle=0]{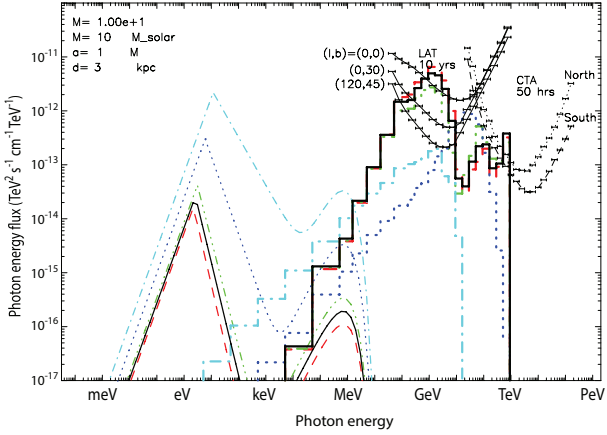}
	\caption{
		\small
		SED of gap emission along the rotation axis, $\theta=0^\circ$,
		for $M=10 M_\odot$ and $a=0.9999M$.
		The distance is assumed to be 3~kpc.
		The cyan dash-dotted, blue dotted, green dash-dot-dot-dotted,
		black solid, and red dashed lines
		correspond to the dimension accretion rates,
		$\dot{m}= 10^{-3.00}$, $10^{-3.50}$, $10^{-4.00}$, 
		$10^{-4.125}$, and $10^{-4.25}$, respectively.
		Thus, the black solid line represents the same case as
		the black solid line of Fig.~\ref{fig:SED_1e1_a9999_5th}.
		The three thin curves in the top right part show
		the detection limit with the {\it Fermi}/LAT 
		with ten-year observations
		at galactic longitude $l$ and latitude $b$
		as designated.
		The dashed and dotted curves in the top right part show
		the Cherenkov Telescope Array detection limit
		after a 50-hour observation
		for the southern and northern sources, respectively.
	}
	\label{fig:SED_1e1_a9999_00dg}
\end{figure}

Let us next examine the gap spectrum 
for $\theta=0^\circ$
as a function of the dimensionless accretion rate, $\dot{m}$.
In Fig.~\ref{fig:SED_1e1_a9999_00dg}, 
we present the SED of gap emission 
for five discrete values of $\dot{m}$, between $10^{-3}$ and $5.6 \times 10^{-5}$.
For $\dot{m} > 10^{-3}$, we cannot solve the gap accurately, because its longitudinal width becomes too small. 
On the other hand, for $\dot{m} < 5.6 \times 10^{-5}$, the vacuum gap becomes non-stationary and we cannot consider this case in the present stationary analysis.
The thin curves on the left show the input ADAF spectra, 
while the thick lines on the right show the output spectra from the gap.
We find that the gap-emitted $\gamma$-ray flux increases 
with decreasing $\dot{m}$, 
because the potential drop in the gap increases with decreasing $\dot{m}$.

The spectra peak between 0.1~GeV and 2~GeV for such stellar-mass BHs,
because the curvature process dominates the IC one.
It is clear that the HE flux lies ten times above
the {\it Fermi}/Large Area Telescope ({\it Fermi}/LAT) detection limit
(the three thin solid curves labeled with ``LAT 10 yrs'')
\footnote{https://www.slac.stanford.edu/exp/glast/groups/canda/\\
	lat\_Performance.htm},
when the accretion rate is 
$5.6 \times 10^{-5} \le \dot{m} \le 1 \times 10^{-4}$.
Consequently, if a nearby, stellar-mass, extreme Kerr BH 
($M \approx 10 M_\odot$, $a \approx 0.9999M$, $d<3$~kpc)
spends more than 10\% of its time in flaring states
(as the green dash-dot-dot-dotted, black solid, and red dashed lines show),
the gap emission will be detectable with the {\it Fermi}/LAT, since the time-averaged flux of the gap emission will appear at ten times lower than the plotted value.
Alternatively, if we collect the photons for 1.2 months, the LAT detection limits will appear at ten times higher value than what are plotted with the thin curves, which means that we can assume a data compilation period of ten years. Therefore, if a BH-gap flare lasts for 1.2 months in a row and if we collect the photons during the same period, we can detect the HE flare with the {\it Fermi}/LAT.

\begin{figure}
		\vspace{-0.4 truecm}
	\includegraphics[width=8.0truecm,angle=0]{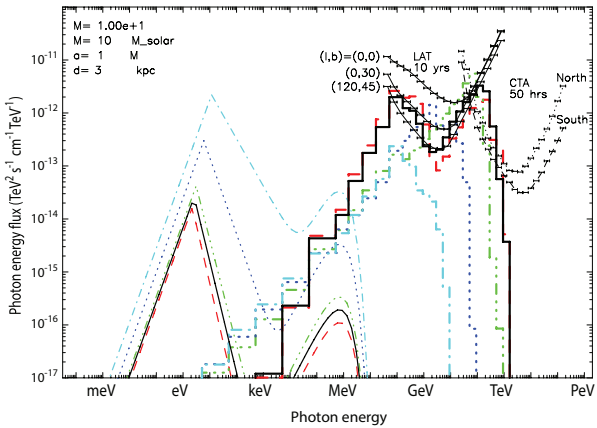}
	\caption{
		\small
		SED of gap emission along a middle latitude, $\theta=45^\circ$,
		for $M=10 M_\odot$ and $a=0.9999M$.
		The distance is assumed to be 3~kpc.
		The lines correspond to the same $\dot{m}$'s as 
		in Fig.~\ref{fig:SED_1e1_a9999_00dg}.
	}
	\label{fig:SED_1e1_a9999_45dg}
\end{figure}

In order to examine the viewing-angle dependence, 
in Fig.~\ref{fig:SED_1e1_a9999_45dg},
we present the SED along $\theta=45^\circ$
for the same case as Fig.~\ref{fig:SED_1e1_a9999_00dg}.
It follows that the VHE flux appears a few times above 
the Cherenkov Telescope Array (CTA) detection limit
(the dashed and dotted curves labeled with ``CTA 50 hrs'')
\footnote{https://portal.cta-observatory.org/CTA\_Observatory/performance/\\
	SitePages/Home.aspx},
when $5.6 \times 10^{-5} \le \dot{m} \le 1 \times 10^{-4}$.
Hence, if a nearby, stellar-mass, extreme Kerr BH 
($M \approx 10 M_\odot$, $a \approx 0.9999M$, $d<3$~kpc)
experiences a VHE flare
(as the green dash-dot-dot-dotted, black solid, and red dashed lines show), and if we view the source at $\theta \approx 45^\circ$,
the gap emission may be marginally detectable with a one-night observation by using the CTA.

\begin{figure}
			\vspace{-0.6 truecm}
	\includegraphics[width=8.0truecm,angle=0]{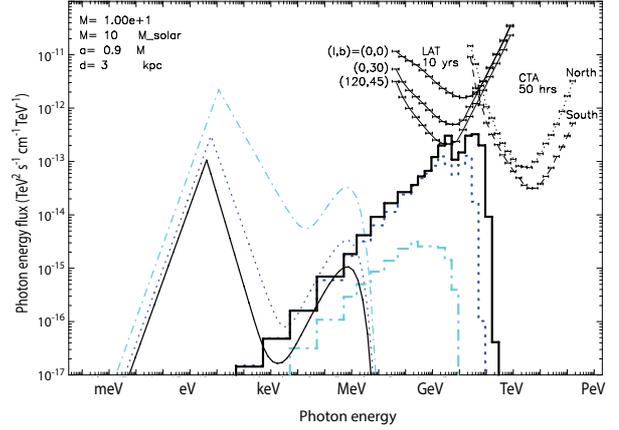}
	\caption{
		\small
		SED of gap emission along the rotation axis, $\theta=0^\circ$,
		for $M=10 M_\odot$ and $a=0.90M$.
		The distance is assumed to be 3~kpc.
		The cyan dash-dotted, blue dotted, and black solid lines
		corresponds to the dimension accretion rates,
		$\dot{m}= 10^{-3.00}$, $10^{-3.50}$, and $10^{-3.75}$, respectively.
	}
	\label{fig:SED_1e1_a9000}
\end{figure}

Now, let us consider a slower spin, $a=0.90M$.
In Fig.~\ref{fig:SED_1e1_a9000}, 
we present the SED along $\theta=0^\circ$.
Since solutions can be found only for $\dot{m} \ge 1.77 \times 10^{-4}$,
we only plot the cases of 
$\dot{m}=10^{-3.0}$, $10^{-3.5}$, and $10^{-3.75}$
as the cyan dash-dotted, blue dotted, and black solid lines.
Comparing with the $a=0.9999M$ case 
(i.e., Fig.~\ref{fig:SED_1e1_a9999_00dg}),
we find that the $\gamma$-ray flux decreases 
more than an order of magnitude.
However, if $B^r$ were constant on the horizon, as assumed in HP16, H16, and H17, the results still do not change very much between $a = 0.90 M$ and $a = 0.9999M$.
As a consequence, we can conclude that the poleward enhancement 
of $B^r$ as $a \rightarrow M$
plays a pivotal role
in the prominent increase of the BH-gap $\gamma$-radiation
for an extremely rotating Kerr BH.

\section{Summary and discussion}
\label{sec:disc}

To summarise, we investigated 
how the BH gap emission is influenced 
by the lateral concentration of the magnetic field lines 
toward the rotation axis, $\theta=0^\circ$, 
as the BH spin approaches its maximum value, $a \rightarrow M$.
For a stellar-mass BH, the gap shows an enhanced
high-energy $\gamma$-radiation along the rotation axis,
when the dimensionless accretion becomes
$5 \times 10^{-5} < \dot{m} < 10^{-4}$.
If a nearby, extremely rotating, stellar-mass BH 
experiences such a flare,
its gap emission will be detectable with the {\it Fermi}/LAT,
provided that we view the BH nearly along the rotation axis.

Actually, it is not very clear  that
if an astrophysical BH can be spun up to 
the near-maximum value, $a=0.9999M$.
For instance, such a BH may preferentially
capture negative angular momentum photons 
radiated from the accreting plasmas
to spin down to the canonical value, $a=0.998M$
\citep{bard70,thorne74}.
In this letter, 
we adopted such a large value as $a=0.9999M$
in order to demonstrate the extreme case.
More moderate cases (e.g., $a=0.99M$)
can be qualitatively interpolated from 
the two cases we have considered, $a=0.90M$ and $a=0.9999M$.
If $a \approx M$,
positive-angular-momentum plasmas will not accrete onto the horizon,
whatever $\dot{m}$ may be small.
Nevertheless, the plasmas will be ejected from the equatorial
ergosphere as an outflow without penetrating in the polar regions,
$\theta \ll 1$.
Therefore, 
in the present argument, it is not essential as to whether or not the plasmas plunge onto the horizon; it is necessary only to form the horizon-penetrating magnetic field.

The poleward enhancement of the $\gamma$-ray flux as $a \rightarrow M$, indeed, 
is unaltered if we adopt different BH masses.
This is because the magnetic field lines concentrate 
toward the rotation axis as $a \rightarrow M$, 
irrespective of the BH mass
\citep{tchek10}.
However, as the BH mass increases, 
the IC process dominates the curvature process,
leading to a poleward enhancement of VHE radiation for SMBHs,
which is to be investigated in a separate paper.

\section*{ACKNOWLEDGEMENTS}

One of the authors (K. H.) is indebted to 
Dr. T.~Y. Saito for valuable discussion on the CTA sensitivity.
This work is supported by the Theoretical Institute for Advanced Research 
in Astrophysics (TIARA) operated under Academia Sinica,
and by the Ministry of Science and Technology of 
the Republic of China (Taiwan) through grants 
103-2628-M-007-003-MY3, 105-2112-M-007-033-MY2, 105-2112-M-007-002, 
103-2112-M-001-032-MY3. 










\bsp	
\label{lastpage}
\end{document}